\documentclass[12pt,preprint]{aastex}

\def\spose#1{\hbox to 0pt{#1\h\> \> SS}}
\def\simlt{\mathrel{\spose{\lower 3pt\hbox{$\mathchar"218$}}
     \raise 2.0pt\hbox{$\mathchar"13C$}}}
\def\simgt{\mathrel{\spose{\lower 3pt\hbox{$\mathchar"218$}}
     \raise 2.0pt\hbox{$\mathchar"13E$}}}

\def\mic{{$\mu$m}}

\def\h2o{H$_2$O}

\def\aple{$\mathrel{\hbox{\rlap{\hbox{\lower4pt\hbox{$\sim$}}}\hbox{$<$}}}$}

\def\apge{$\mathrel{\hbox{\rlap{\hbox{\lower4pt\hbox{$\sim$}}}\hbox{$>$}}}$}

\def\g45{G45.45$+$0.6}

\slugcomment{}
\lefthead{Blum {\it et al.}}
\righthead{SAGE Evolved Stars}

\begin{document}
\title{The Ionizing Stars of the Galactic Ultra--Compact \ion{H}{2} Region
G45.45$+$0.06}

\author{Robert D. Blum\footnote{NOAO Gemini Science Center, 950 North
Cherry Avenue, Tucson, Arizona, 85719} }

\author{Peter J. McGregor\footnote{The Australian National University,
Cotter Road, Weston Creek, ACT 2611, Australia.} }

\begin{abstract}

Using the NIFS near--infrared integral--field spectrograph behind the
facility adaptive optics module, ALTAIR, on Gemini North, we have
identified several massive O--type stars that are responsible for the
ionization of the Galactic Ultra--Compact \ion{H}{2} region
G45.45$+$0.06. The sources ``m'' and ``n'' from the imaging study of
\citet{feldt98} are classified as hot, massive O--type stars based on 
their $K-$band spectra. Other bright point
sources show red and/or nebular spectra and one appears to have cool
star features that we suggest are due to a young, low--mass pre--main
sequence component.  Still two other embedded sources (``k'' and ``o''
from Feldt et al.) exhibit CO bandhead emission that may arise in
circumstellar disks which are possibly still accreting. Finally,
nebular lines previously identified only in higher excitation
planetary nebulae and associated with \ion{Kr}{3} and \ion{Se}{4} ions
are detected in \g45.

\end{abstract}

\keywords{infrared: stars, instrumentation: adaptive optics, (ISM:)
\ion{H}{2} regions, (ISM:) dust, extinction, stars: formation,} {\it
Facilities:} \facility{Gemini:Gillett ()}

\section{Introduction}

G45.45$+$0.06 is a luminous ultra--compact \ion{H}{2} (UC\ion{H}{2})
region identified by \citet{wc89} as a ``cometary'' type in their
landmark radio survey. UC\ion{H}{2} regions have become synonymous
with the earliest phases of massive star birth \citep{church02} and
the phase in which the object is first revealed as an identifiable
hot, massive star as evidenced by its strong ionizing
radiation. \citet{wc89} noted that many of the UC\ion{H}{2} regions
must be excited by multiple sources, in effect harboring clusters of
stars. This has indeed turned out to be the case. \citet{feldt98}
presented high angular resolution near infrared images of
G45.45$+$0.06 that showed multiple point sources and a complex
morphology.

We have used the new Near--infrared Integral--Field Spectrograph
(NIFS) on the Gemini North Fredrick C. Gillett 8--m telescope to carry
out a detailed investigation of this UC\ion{H}{2} region and build on the
\citet{feldt98} imaging study. UC\ion{H}{2} regions typically have sizes
\aple 0.1 pc. At the distance of G45.45$+$0.06 \citep[6.6
kpc]{feldt98} this scale just fills the NIFS three arc second
field. Using the facility adaptive optics module ALTAIR with NIFS, we
obtained $\sim$ 0.1$''$ images in several pointings in the larger
\citet{feldt98} field. Adaptive optics on large aperture telescopes
offers a significant advance in the study of massive stars in their
earliest phases since the high angular resolution improves the point
source contrast against the strong and variable nebular
background. NIFS has the added benefit of a new generation
2048$\times$2048 HgCdTe array that provides broad spectral coverage
(an entire near--infrared band) at medium spectral resolution ($R=$
5000).

Previous studies of UC\ion{H}{2} region central ionizing sources in the
near--infrared have identified hot stars \citep{wat97, han02, bik05}
by their photospheric features. These studies provide important
constraints on the statistics of the youngest phase of massive stellar
populations. Coupling such studies to integral--field observations will
advance our understanding of the embedded stellar sources as well, but
the added high angular resolution imaging dimension will allow for
detailed modeling of ionized (and molecular) gas as well as the
clustering characteristics on the smallest scales. \citet{mess07} have
presented results for the Galactic UC\ion{H}{2} region [DBS2003]8 based on
similar observations to those described here and made with the ESO VLT
SINFONI integral--field unit spectrograph.

\section{Observations}

Data were obtained with NIFS at the Cassegrain focus of the Gemini
North Fredrick C. Gillett 8--m telescope on Mauna Kea, Hawaii on the
nights of 22 and 23 July, 2006 (local time). NIFS was used with the
facility adaptive optics (AO) module ALTAIR
\footnote{http://www.gemini.edu/sciops/instruments/altair/altairIndex.html}
in natural guide star (NGS) mode.

NIFS slices an approximately three arc second by three arc second field
into 29 spectral segments of 0.1$''$ width (the ``slit'' width) and
$\sim$ three arc seconds in length.The scale along the spatial
dimension is 0.043$''$ pix$^{-1}$. The resulting ``spaxels'' are thus
0.043$''$ $\times$ 0.1$''$, and each contains a full spectrum covering
one of the near--infrared bands. See \citet{pm03} for a complete
description of NIFS.

In the present case, $K-$band spectra are presented for G45.45$+$0.6
for two pointings from the field described by \cite{feldt98}. The two
pointings were chosen to include candidate ionizing sources identified
by \cite{feldt98}.The AO guide star used for ALTAIR is
located about 12$''$ to the north and east from the two NIFS
pointings. The AO guide star has an $R$ magnitude of 11.8 according to
the USNO catalog (source id: U0975\_14397240). The conditions were
sufficient with the AO correction to isolate all the point sources
identified by \cite{feldt98} within the NIFS field of view. On the
22nd of July, the sky was clear, and the seeing (at 5000 \AA) during
the observations (pointing 1) was approximately 0.6$''--0.8''$, and
the observations were obtained at relatively high airmass
(1.6--1.9). ALTAIR was run at 500 Hz for these observations. On July
23rd, there were thin clouds and the seeing was again about 0.6$'' --
0.8''$ but this time for an observation airmass near 1.0 (pointing 2).
ALTAIR was run at 100 Hz for these observations due to the clouds.

Each pointing consisted of a set of undithered frames taken on source
as well as a sky frame obtained on a nearby ($\sim$ 50$''$ West) blank
field. The pointings had total exposure times of 1800 seconds
(3$\times$600s) on source and 600 seconds on sky. At the time of the
observations, shortly after NIFS commissioning, it was judged best to
stare on source to simplify combining of the frames in the
post--observation reduction. NIFS has a 2048$\times$2048 HgCdTe
detector (HAWAII--2RG) with a significant number of hot pixels that
were removed by sky subtraction. Current NIFS observations are best
taken with small dithers to help reduce the effect of the hot
pixels. The relatively long exposures used to reach
near--sky--noise--limited performance also result in significant
numbers of cosmic rays. These can be mitigated by multiple exposures
as well. The NIFS dark current is quite low, apart from hot pixels,
and is measured to be \aple 0.1 e$^-$ s$^{-1}$. The RMS read noise is
approximately six e$^{-}$ in low noise mode which uses multiple reads
of the detector.

The spectral resolving power of NIFS in the $K-$band is $\lambda /
\Delta\lambda =$ 5160 which results in a linear dispersion of 2.13
\AA/pixel. This dispersion, combined with the large format array gives
a full wavelength coverage at $K$ of about 4200 \AA \ accounting for
some minor truncation in the final data cube due to the systematic
shift of wavelength in each slitlet from the staircase design of the
image slicer.

\section{Data Reduction}

The line maps and spectra presented here were obtained using the
Gemini NIFS IRAF\footnote{IRAF is distributed by the National
Optical Astronomy Observatories, which are operated by the Association
of Universities for Research in Astronomy, Inc., under cooperative
agreement with the National Science Foundation.} data reduction
package (version 1.9).

The NIFS IRAF package allows for full reduction to the ``image cube''
stage where a finalcube has a roughly 60$\times$62 pixel image plane
and a 2040 pixel spectral depth. First, raw images are prepared for
reduction by standard Gemini procedures that create the FITS image
variance and data quality extensions. Next, the data are sky
subtracted, flatfielded, rectified spatially, and wavelength
calibrated. The last two steps result in a uniform spatial--spectral
trace for each image slice row and a linear wavelength scale. It is
important to remember that the final IRAF data cube resamples each
slitlet to two 0.05$''$ pixels for convenience; the angular resolution
in this dimension is still 0.1$''$. It might be possible to enhance
the resolution in the slitlet spatial dimmension by careful dithering
of the telescope and combining of the data, but this is not the
default situation for NIFS that provides native 0.048$''$ $\times$
0.1$''$ ``pixels'' on the sky.

The spatial rectification was accomplished by use of a Ronchi mask
image taken as part of the baseline calibration set for NIFS.  The
Ronchi mask is a coarse transmission grating deployed in the
instrument focal plane and illuminated by the flatfield calibration
lamp. It produces a uniform distribution of compact
artifical sources along each image slice. In the spatial dimension,
the Ronchi mask produces nine spectral traces for each slice. Each
trace can be thought of as the trace of a point source at a given
spatial position of each slice. Fitting the traces for all slices
primarily removes a small systematic ($\sim$ $\pm$ one pixel
end--to--end) slope of the spectral trace and allows for an accurate
reconstruction of the 2--D image since the traces of one slice can be
identified with the traces of neighboring slices.

A lamp image of Ar and Xe lines was used to determine the dispersion
along each slice and as a function of the spatial dimension of each
slice. A quadratic polynomial was used in this case which
produced typical uncertainties in the position of a spectral line of
approximately $\pm$ 0.2 \AA \ ($\sim$ 1/10 pixel). Final spectra are
interpolated to a linear wavelength solution.

The spectra were next corrected for telluric absorption by division by
the spectrum of an A0 V star. Each A0 V (in this case HIP84147 and
HIP89309 on July 22 and HIP99796 on July 23) was corrected for
intrinsic Br$\gamma$ absorption by fitting a Voigt profile to the
telluric spectrum. This approximate correction is sufficient for our
purposes since the hot star photospheric classifications do not rely
on Br$\gamma$. 

The fully rectified and telluric corrected images were used to extract
point sources using the NFEXTRACT routine. In the Gemini IRAF NIFS
package, the point source extraction is done on the image that still
contains the 29 slices, i.e. before construction of a final data cube
image (the extraction aperture is applied to a temporary cube).  In
this case, the NFEXTRACT routine was modified to allow for a secondary
sky spectrum to be subtracted from the point source to account for
over-- or under--subtraction of sky between the on--source and sky
images. In both G45.45$+$0.06 pointings, a one arc second diameter aperture was
used to measure the residual sky emission. This aperture obviously
includes some nebular emission, but the sky aperture location was
chosen to minimize this effect. For pointing 1, this was a location to
the South and West of the point sources near the edge of the image
(see below). For pointing 2, the sky aperture was located to the North
and West of the point sources near the edge of the image. All point
sources were extracted using a 0.4$''$ diameter aperture. This
aperture was chosen to minimize the overlap between point sources
while attempting to obtain as much flux as possible and is
distinguished from the larger aperture used for flux calibration (see
next) that was chosen to match the photometry of \citet{feldt98}.

The final wavelength calibrated, spatially rectified, and telluric
corrected images were transformed into data cubes. The spatial pixel
scale was interpolated to 0.05$''$ in the fine dimension and block
replicated to 0.05$''$ in the course dimension providing for a uniform
scale as described in the previous section. Line maps presented below
were extracted from these cubes using NFMAP. A rough flux calibration
was done by extracting spectra in 0.5$''$ apertures and comparing them
to the photometry of \citet{feldt98} from the same diameter aperture
(source ``o'' in pointing 1, and sources ``g'', ''h'', ''i'', and
``k'' in pointing two). No attempt was made to correct the fraction of
light outside this aperture in the seeing halo for this approximate
flux calibration.  Making an aperture correction on the science frames
is not possible because of the strong nebular emission; however, the
standard stars taken before and after the science observations (under
similar seeing as the science targets) suggest that a 0.5$''$ diameter
aperture misses at least 25$\%$ of the stellar flux. The following
results and discussion do not depend strongly on the absolute
calibration of the NIFS data. This calibration applies only to the
extended emission maps presented below; point sources are calibrated
directly to the \citet{feldt98} $K'$ magnitudes
($\lambda_{\circ}=$2.12 \mic) using a flux zero point of
4.75$\times$10$^{-11}$ erg cm$^{-2}$-s$^{-1}$-\AA$^{-1}$. This gives
$Ks$ ($\approx$ $K$) $=$ 12.4 and 12.0 mag for sources ``m'' and
``n'', respectively.

\section{Results}

\begin{figure}
\plotone{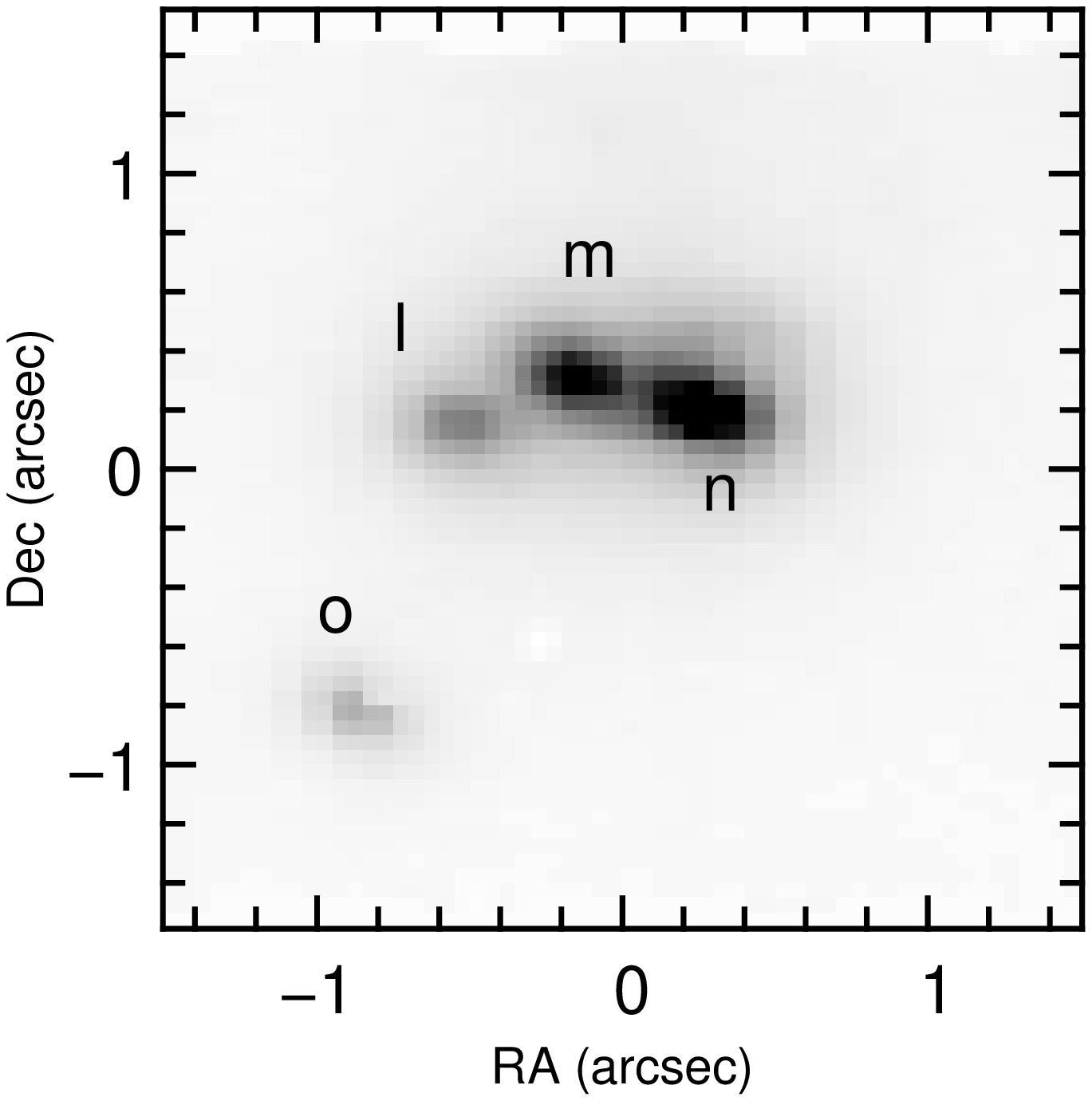} 
\caption{Continuum map at 21695 \AA \ (10.7 \AA \ wide) for the first
NIFS pointing.  The point sources are labeled following
\cite{feldt98}. Sources m and n exhibit spectra of mid O--type stars
while source ``o'' has CO band head emission at 2.3 \mic \ suggesting it
may be a massive young stellar object; see text and Figures
\ref{otype} and \ref{co}. The field center is located at approximately
RA(2000)$=$19$h$~14$m$~21.24$s$,
Dec(2000)$=+$11$d$~09$'$~10.1$''$. North is up and East to the left in
the image, as indicated.
\label{217a}}
\end{figure}

\begin{figure}
\plotone{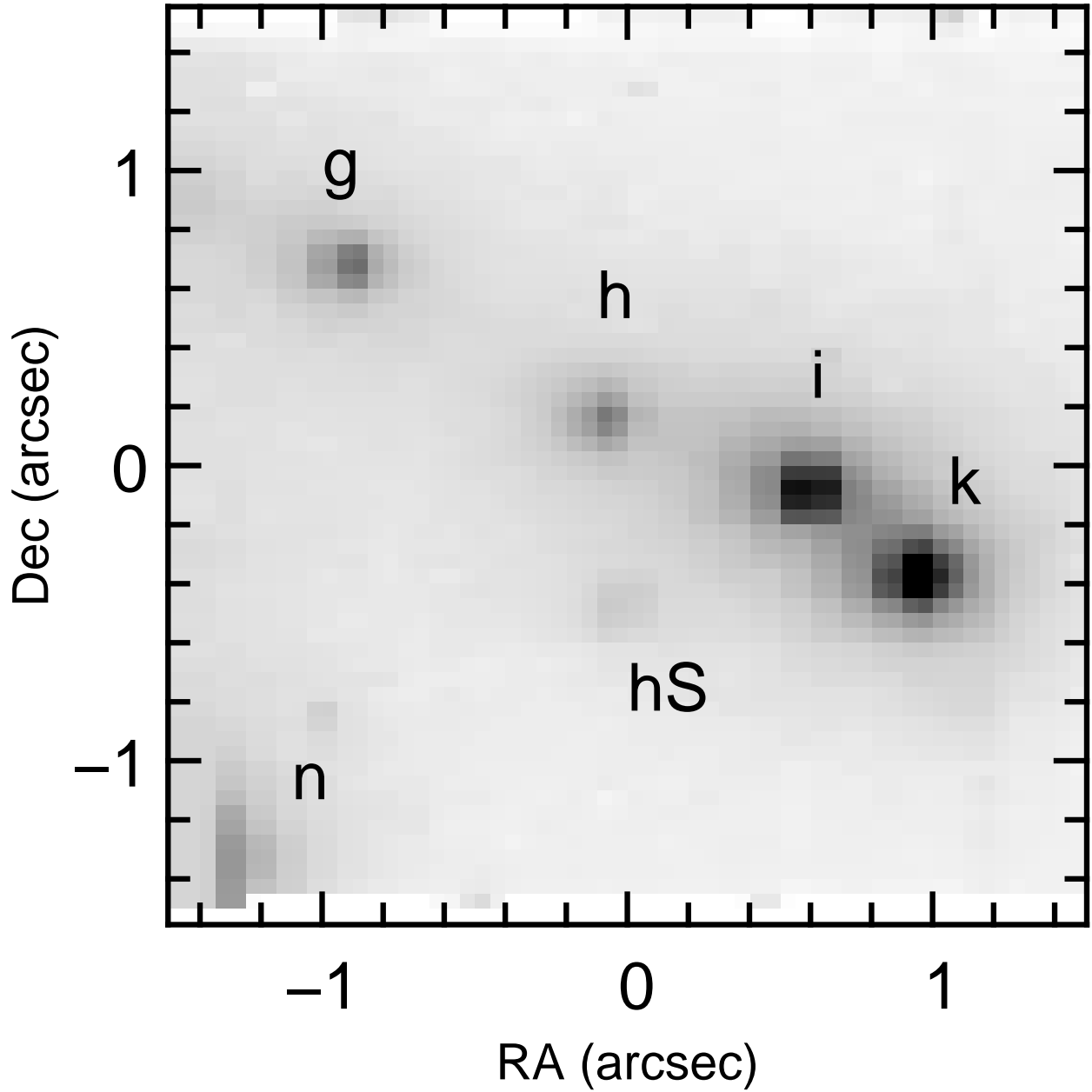}
\caption{Continuum map at 21695 \AA \ (10.7 \AA \ wide) for the second
  NIFS pointing (Figure~\ref{217b}). The point sources are labeled
  following \cite{feldt98}; ``hS'' is newly identified in this
  work. Sources ``g'', ''h'' and ''i'' exhibit spectra with
  featureless red continua and source ''k'' shows CO band head
  emission similar to source ''o''; see text and Figures \ref{neb} and
  \ref{co}.  The field center is located at approximately
  RA(2000)$=$19$h$~14$m$~21.14$s$,
  Dec(2000)$=+$11$d$~09$'$~11.9$''$. North is up and East to the left
  in the image, as indicated.
\label{217b}}
\end{figure}

Figures~\ref{217a} and \ref{217b} show continuum emission for
pointings one and two, respectively, extracted from the final
data cubes. The wavelength for the extracted band is 21695 \AA, and the
band is approximately 10.7 \AA \ wide. The point sources in these images
have image cores of about 0.15$''$-0.25$''$. This is for a final set
of three combined images. The point source designations of \cite{feldt98}
are indicated in each figure, and all of them are clearly and easily
separated, even though they are deeply embedded in strong nebular
emission from the UC\ion{H}{2} region. A faint source not identified by
\cite{feldt98} appears about 0.5$''$ south of source ``h'' in
Figure~\ref{217b}; we designate this source ``hS''.

\begin{figure}
\plotone{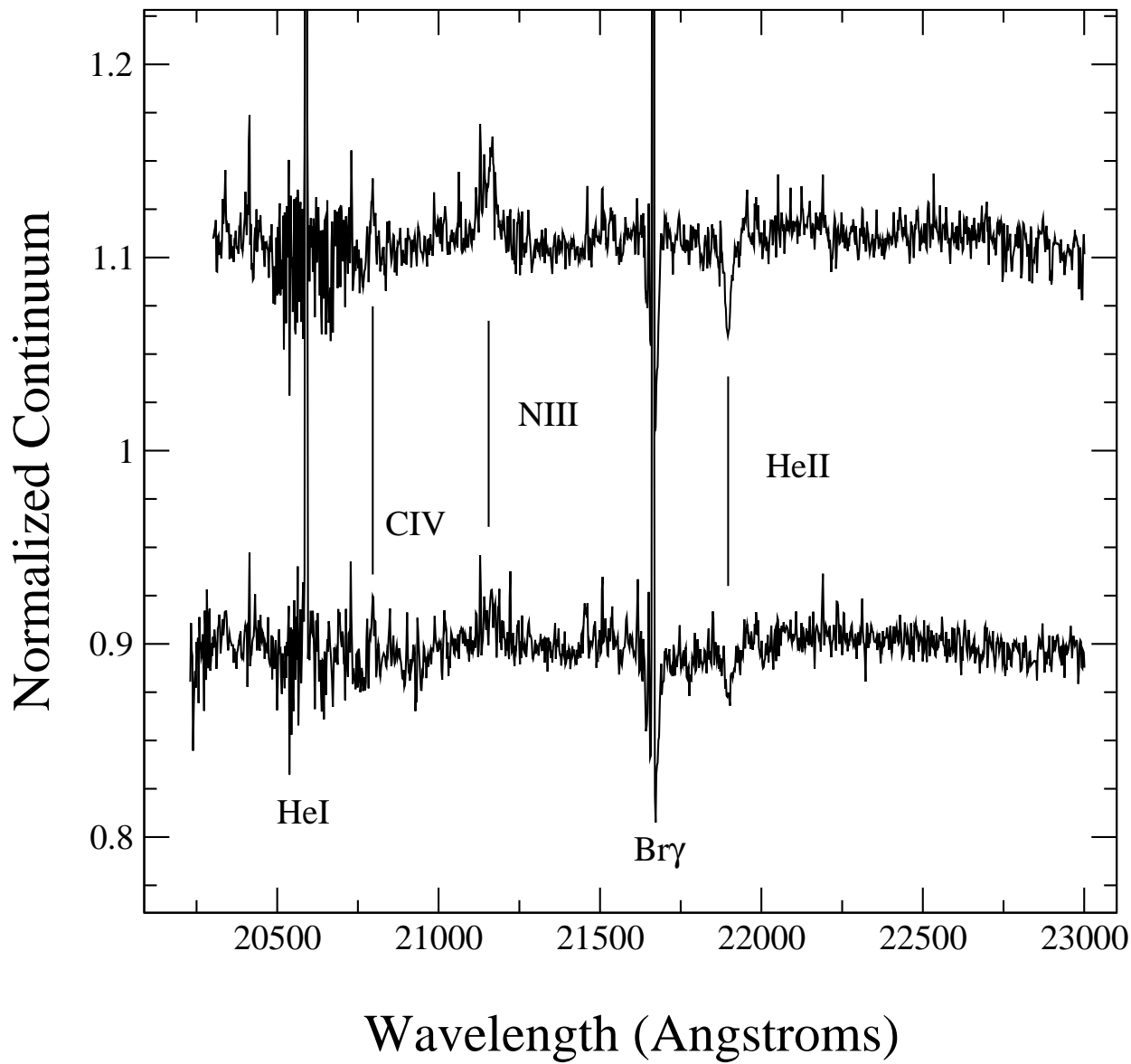}
\caption{Spectra of O--type stars, ``m'' and ``n,'' from the first
  NIFS pointing; see Figure~\ref{217a}. Lines of \ion{C}{4},
  \ion{He}{2}, and \ion{N}{3} identify these sources as mid or
  earlyO--type; see text. The spectra have been arbitrarily restricted
  in wavelength to show the weak stellar features. The spectral
  resolution is $R (\lambda/\Delta\lambda) = 5160$. These spectra were
  normalized by a low order fit to the continuum after division by an
  A0 V type standard.
\label{otype}}
\end{figure}

\begin{figure}
\plotone{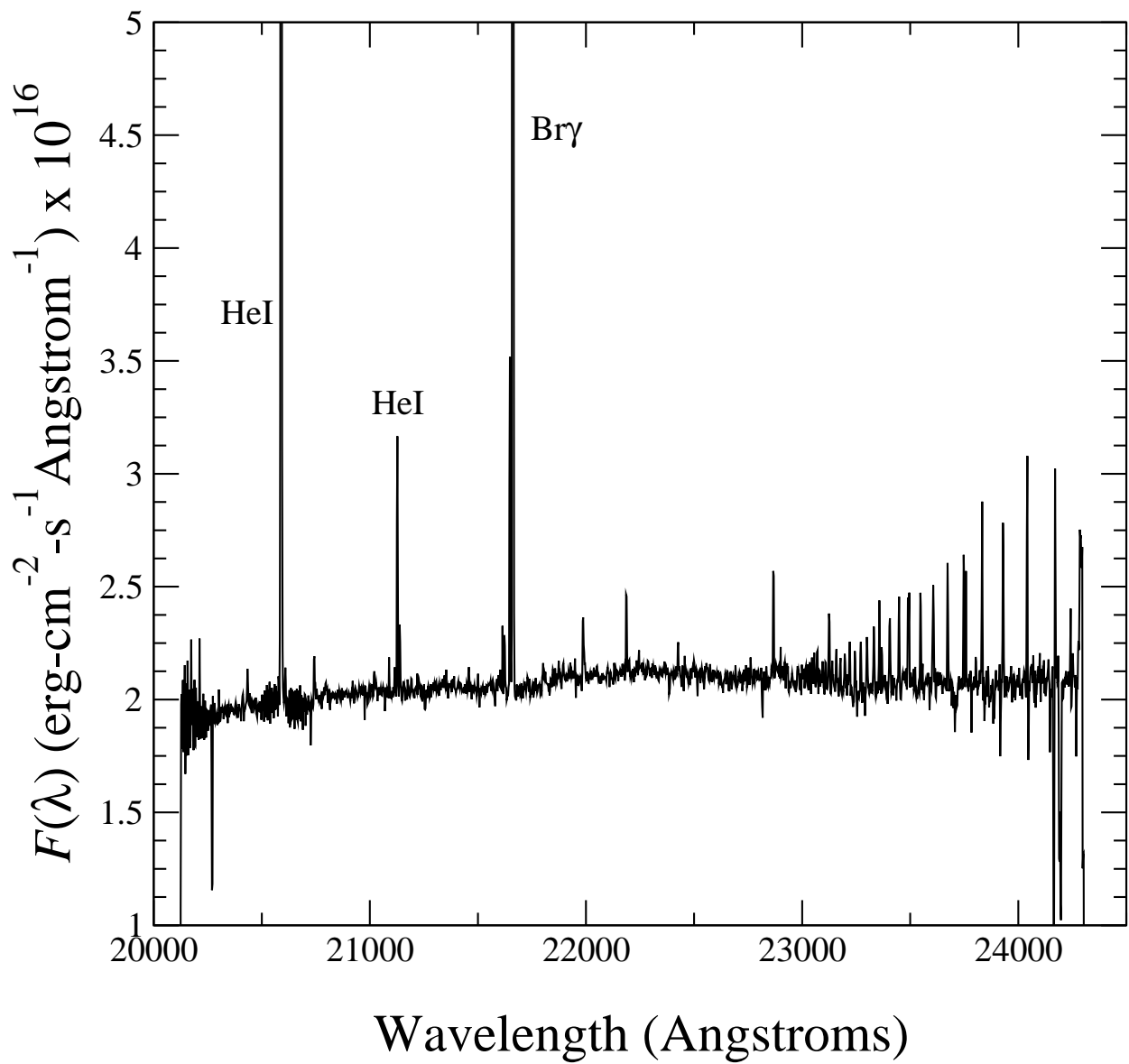}
\caption{Combined spectrum of stars with nebular features, but no hot
star photospheric features: ``l'', from the first NIFS pointing and
``g'', ``h'', ``i'' from the second; see Figures~\ref{217a} and
\ref{217b}.  The spectral resolution is $R (\lambda/\Delta\lambda) =
5160$. These spectra were normalized by an A0 V type standard and
multiplied by a black body spectrum corresponding to $T_{\rm
eff}=9000$K. The flux scale was set by assigning the flux at 2.12 \mic
\ according to the magnitudes given by \citet{feldt98} and a flux zero
point at 2.12 \mic \ of 4.75 $\times$10$^{-11}$
erg s$^{-1}$-cm$^{-2}$-\mic$^{-1}$. Typical nebular lines of H and He
are shown; other emission lines are described in the text and
Figure~\ref{cozoom}.
\label{neb}}
\end{figure}

\begin{figure}
\plotone{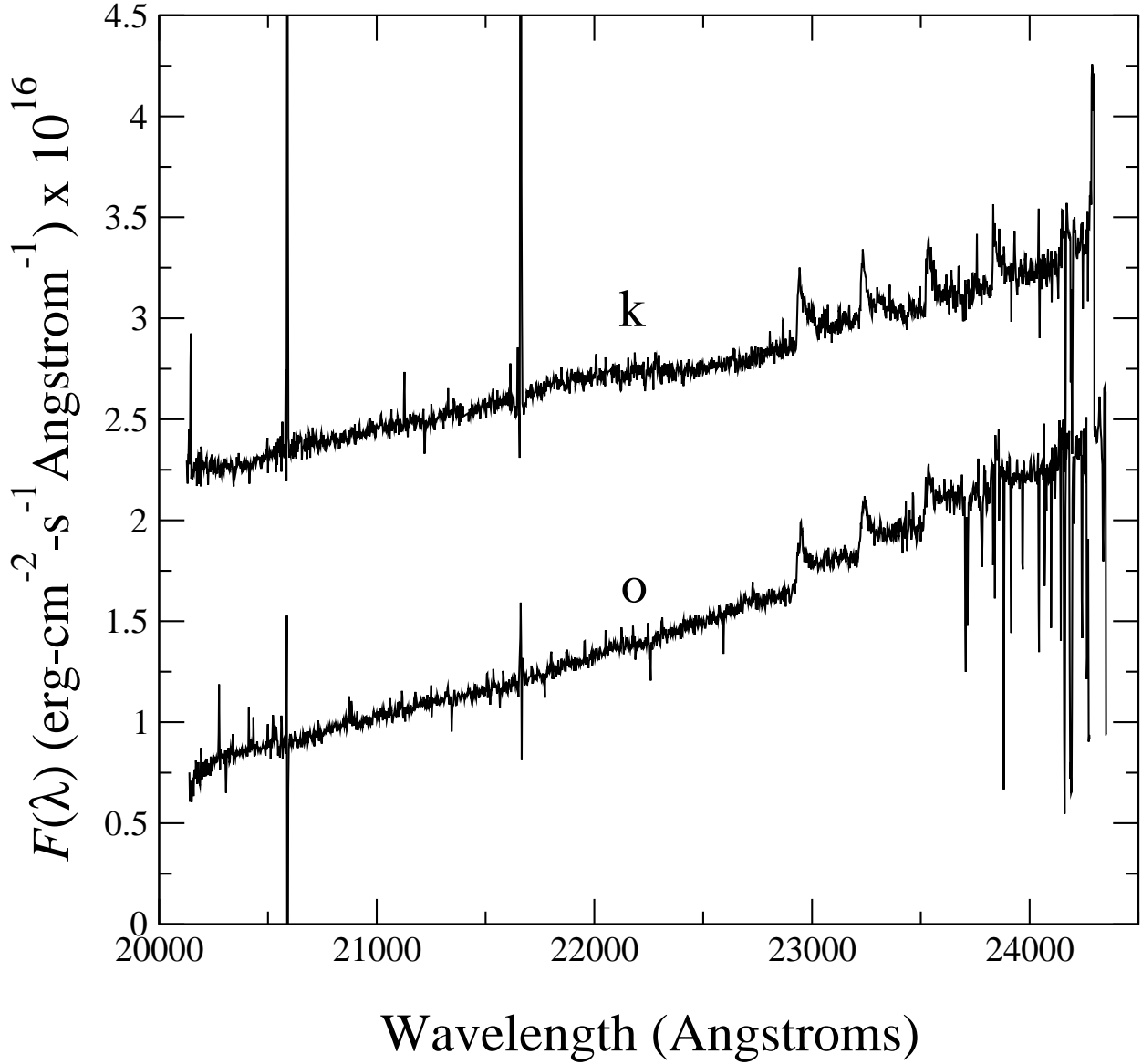}
\caption{Spectra of CO emission stars,o, from the first NIFS pointing
and k from the second; see Figures~\ref{217a} and \ref{217b}.
The spectral resolution is $R (\lambda/\Delta\lambda) = 5160$. These
spectra were normalized by an A0 V type standard and multiplied by a
black body spectrum corresponding to $T_{\rm eff}=9000$K.
\label{co}}
\end{figure}

Point source spectra are shown in
Figures~\ref{otype}--\ref{co}. \citet{feldt98} sources ``m'' and ``n''
exhibit photospheric features of mid--early O--type stars
\citep{hcr96}.  Though the telluric correction is relatively poor near
20600 \AA, there appears to be weak \ion{C}{4} emission at 20796 \AA
\ and perhaps at 20842 \AA \ in both sources \citep[vacuum line positions 
from][]{nist}. The former
line is stronger, as is typical \citep{hcr96}. The presence of
\ion{C}{4} emission near 20800 \AA \ together with the strong
\ion{He}{2} [absorption] and \ion{N}{3} [emission] features indicates
a spectral type earlier than about kO6 and later than about
kO4. Sources ``k'' and ``o'' have CO overtone emission starting with
the 2--0 rotational--vibrational bandhead at 22935 \AA. These sources
do not show obvious photospheric features. There is a suggestion of
weak and broad absorption at the postion of Br$\gamma$ in the spectrum
of source ``k;'' however, it is not possible to claim an unambiguous
detection due to the strong nebular emission seen toward source ``k''
and the (imperfect) correction of Br$\gamma$ absorption in the
telluric standard. Apart from CO bandhead emission, sources ``k'' and
''o'' exhibit strong continuum emission and overlying, compact,
nebular emission.

\begin{figure}
\plotone{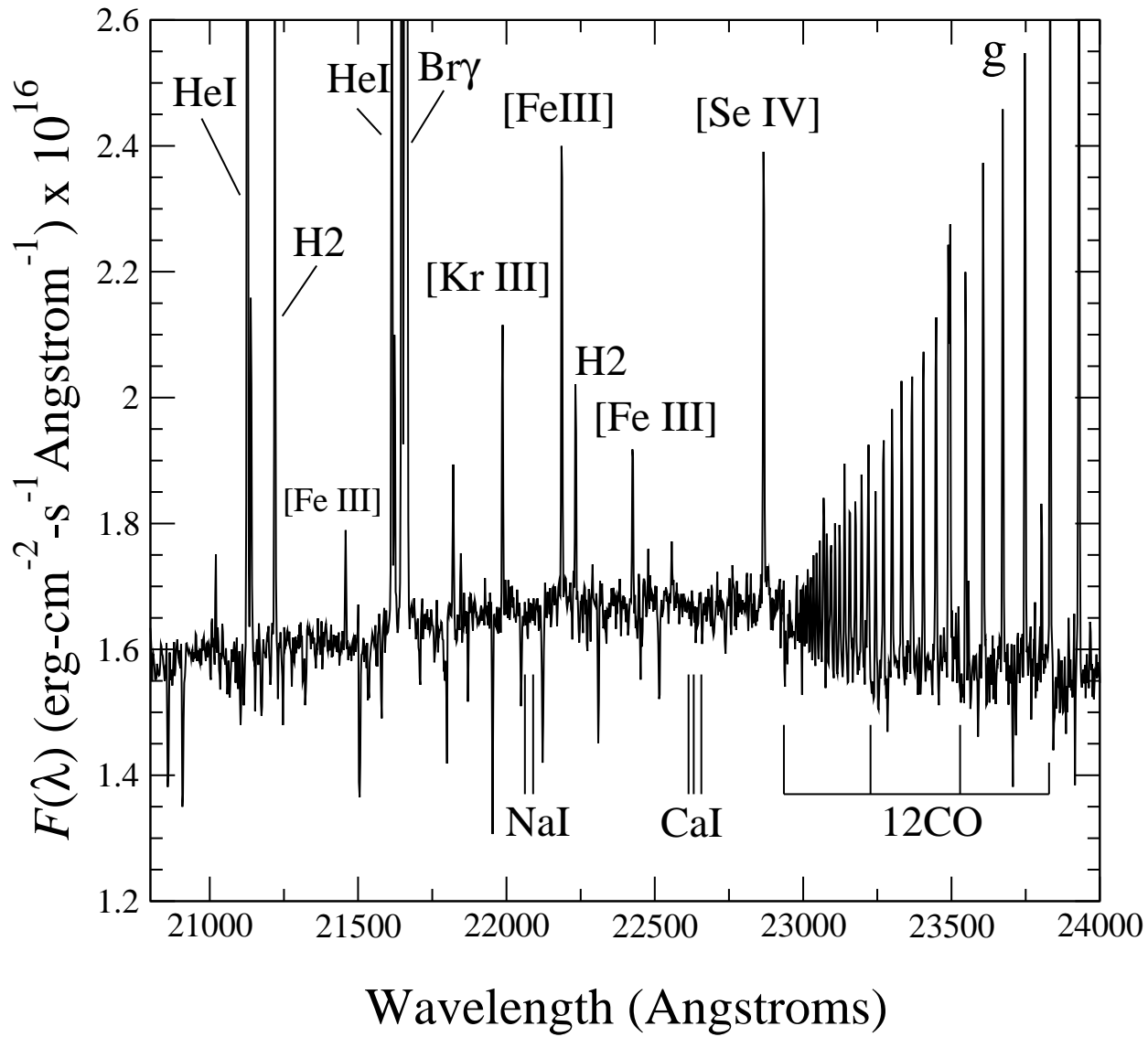}
\caption{Same as Figure~\ref{neb}, but for source ``g'' alone and
zoomed around the region of the CO first overtone bandhead at 22935
\AA. There is an absorption component to the flux of source ``g'' due
to young, lower mass star(s) in the ultra--compact \ion{H}{2} region; see
text. The positions of cool--star features due to NaI and CaI are
indicated, but not confidently detected. The rich H I Pfund series is
seen to the red side of 23000 \AA. Other lines are also indicated (see text
and Figure~\ref{krsel}).
\label{cozoom}}
\end{figure}

Sources ``h'', ``i'', and ''l'' show only nebular features
superimposed upon a red continuum.  Source ``g'' exhibits a strong
nebular spectrum, but upon close inspection, there appear to be ``cool--star''
absorption features superimposed upon the spectrum as
well. Figure~\ref{cozoom} shows the region of source ``g'' around the
region of the CO 2--0 rotational--vibrational bandhead at 22935 \AA.

\begin{figure}
\plotone{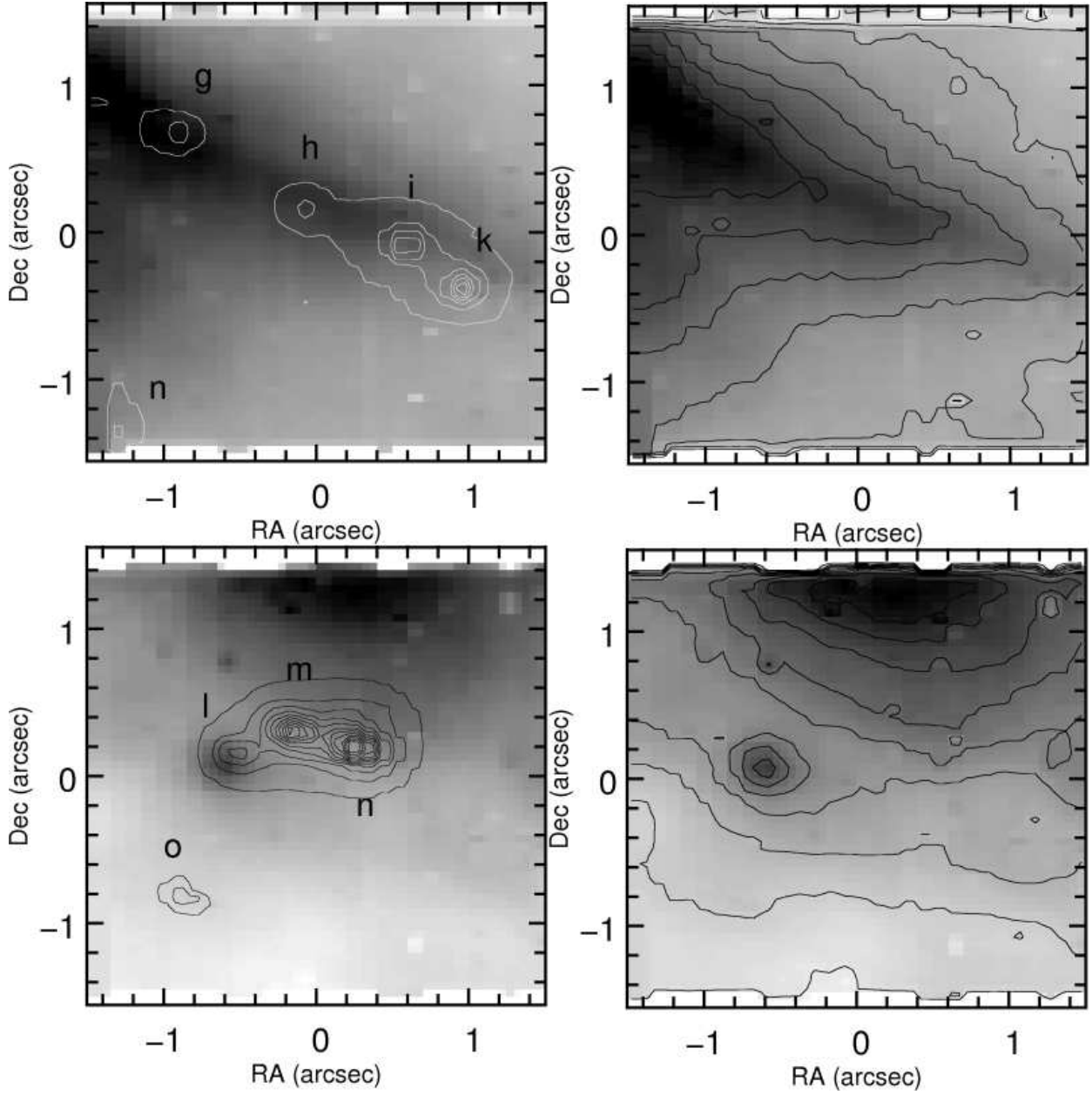}
\caption{Br$\gamma$ line maps for pointing 1 ({\it bottom} row) and
pointing 2 ({\it top} row). The continuum contours for a band at 21700
\AA \ are overlaid on the left panels while the Brg$\gamma$ contours
themselves are overlaid on the right. There is a compact Br$\gamma$
source in the continuum subtracted map for pointing 1 that appears to
be associated with source ``l''. The Br$\gamma$ contour levels (per
0.05$\times$0.05 arc second$^{2}$ pixel) are spaced by
1.1$\times$10$^{-17}$ erg cm$^{-2}$--s$^{-1}$ and
2.4$\times$10$^{-17}$ erg cm$^{-2}$--s$^{-1}$ in pointing 1 and 2,
respectively. The peak contour level is 10 times the contour spacing
in each pointing.
\label{brg}}
\end{figure}

There are strong nebular emission lines throughout the three arc second field
of view in pointings 1 and 2. Br$\gamma$ (21661 \AA) and He I (20587
\AA \ and 21127 \AA) lines are strong as expected for an \ion{H}{2} region
with embedded hot stars. Figure~\ref{brg} shows the distribution of
ionized gas in each pointing along with an overlay of the contours of
the continuum emission. The continuum band is centered near 21700 \AA
\ and it highlights the positions of the point sources relative to the
ionized gas. A ``point source'' appears in the continuum--subtracted
Br$\gamma$ map for pointing 1 that is near to, but not quite
centered on, the position of source ``l''. The ionized gas appears to
form a sharp ridge in pointing two centered on the line of
(presumably) embedded point sources, though the point sources appear
systematically offset to the south as the ridge is traversed from east
to west.

The peak Br$\gamma$ fluxes in each pointing are consistent with the
narrow--band map presented by \citet{feldt98}. The
ratio of He I (21127 \AA) to Br$\gamma$ is roughly uniform over both
pointings and equal to about 0.035 $\pm$ 0.003 (not corrected for
reddening).

\begin{figure}
\plotone{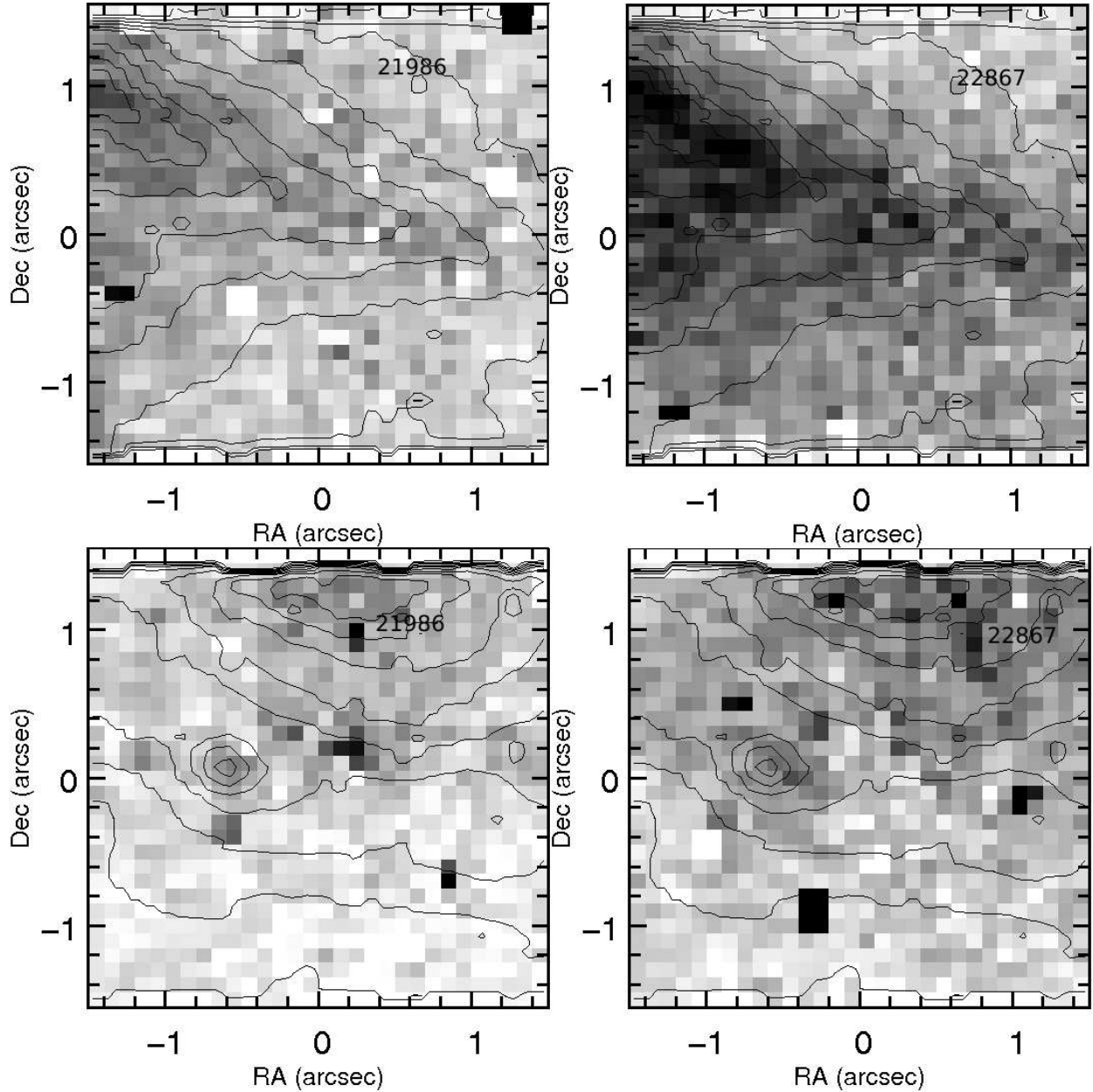}
\caption{Line maps for pointing 1 ({\it bottom} row) and pointing 2
({\it top} row) for emission centered near 21986 \AA \ and 22867
\AA. Br$\gamma$ contours themselves are overlaid on each panel. To
improve signal to noise, these maps have been block averaged to
0.1$''$ by 0.1$''$ pixels (i.e. 2 by 2 binning).
\label{krsel}}
\end{figure}

Two lines present in the data cubes have been identified previously in
planetary nebulae (PNe) spectra. These are the lines at 21986 \AA \ and
22867 \AA \ that were attributed to [\ion{Kr}{3}] and [\ion{Se}{4}],
respectively, in spectra of PNe by
\citet{din01}. Figure~\ref{krsel} shows the gray--scale maps for both
of these lines in each pointing (pointing 1, bottom; pointing 2,
top). Each map is overlaid with the Br$\gamma$ contour, and it
appears that the 21986 \AA \ and 22867 \AA \ lines trace the ionized gas
region. All the lines detected are identified in
Table~1. The line intensities with respect to Br$\gamma$ are
similar in both pointings and all lines in the table were
detected in both pointings.

\section{Discussion}

\citet{feldt98} used adaptive optics imaging at near--infrared
wavelengths ($H$, and $K'$ in conjunction with lower angular
resolution radio and mid--infrared images to describe the evolutionary
state of the UC\ion{H}{2} region and its massive stars. Though their 4--m
telescope data were still of insufficient resolution to make accurate
photometric measurements of the bright trio of sources ``l'', ``m'',
and ``n'', they speculated that these sources were the more massive
ionizing objects embedded in G45.45$+$0.06 and had formed first,
triggering a second star formation event along the ridge of ionizing
emission that gives this UC\ion{H}{2} its original cometary designation
\citep{wc89}. Our NIFS spectra of sources ``m'' and ``n'' confirm the
first part of this scenario, at least. The two sources are hot,
massive stars as identified by their photospheric 
features (Figure~\ref{otype}). G45.45$+$0.06 is thus one of only a few
UC\ion{H}{2} regions with spectroscopically confirmed central ionizing
sources of the hottest spectral types \citep{wat97, han02, bik05},
though identification of central stars in UC\ion{H}{2} regions of mid O, late
O, and early B types is now relatively common \citep{bik05}.

Along with the O--type stars, we have detected CO bandhead emission in
sources ``o'' and ``k''. This feature has been associated with massive
young stellar objects \citep{bik04, blum04, bik06}.  The brightness of
source ``k'' \citep[$K' =$ 14.1; see][]{feldt98} suggests it might be a
hot star if its $H-Ks$ color were the result of reddening
only. However, candidate massive stars with CO emission typically have
strong excess emission due to reprocessing of the central star
radiation by the circumstellar disk and/or envelope. It is more likely
that the CO emission sources we observe in G45.45$+$0.06 are somewhat
lower mass objects than the identified O--type stars and have not yet
dispersed their natal cocoons as in W31 \citep{blum01}. See
\citet{blum04} for an extended discussion of the near--infrared excess
emission observed around young hot stars with CO emission.

The location of the CO emission sources and the apparently embedded
sources ``g'', ``h'', and ``i'' outside the central ``cluster'' of O
stars is consistent with a picture as described by \citet{feldt98} in
which the O stars have triggered further star formation. However, as
mentioned above, such embedded objects usually appear bright at $K$
due to strong excess emission \citep[see the discussion in][]{blum04},
perhaps due to an accretion disk. The expected $K$ excess for objects
with reprocessing disks can be quite large (2--4 mag) with a
corresponding reduction in the resulting mass/luminosity
\citep{hil92}. The disk (or envelope) dissipation time is longer for
lower masses, so the issue of which stars formed when is
unresolved. The strongest argument for triggered star formation
remains the morphology of the sources embedded in the UC\ion{H}{2}
region. The source ``g'' appears to be the most heavily embedded
source and the ionized gas systematically becomes separated from the
point sources along the NE--SW ridge of emission as one moves toward
source ``k'' to the SW (see Figure~\ref{brg}). The impression is that
the hot stars ``m'' and ``n'' to the SE are ionizing this ridge and
that if they are responsible for triggering the star formation events
associated with the embedded sources, then perhaps source ``k'' formed
first and has evolved slightly as the ionization front moves
outward. Alternately, the lower mass stars have formed at the same
time, but around the central high mass stars.

The UC\ion{H}{2} region is powered by more than one massive star, and
there are a number of lower mass hot (presumably B--type) stars as
indicated by the NIFS spectroscopy. The exciting ``star'' of this
UC\ion{H}{2} region is, in fact, a cluster, as pointed out by
\citet{feldt98}; the NIFS data now provide a more quantitative
description of the stellar content. This includes lower mass stars:
Figure~\ref{cozoom} suggests we have detected a cool--star
component. The region of the spectrum near and beyond 23000 \AA
\ shows what appears to be bandhead absorption due to the CO
molecule. The signal is strongest in the aperture around source ``g''
and it is not clear if source ``g'' itself is responsible for all the
absorption or if there are one or more cooler stars very close to an
otherwise more massive star. The strong Pfund series emission
complicates the situation and is also contributing to the appearance
of a lower ``continuum'' beyond 23000 \AA
\ \citep[see][]{kraus00}. Nevertheless, a strong feature appears at
the wavelength of the CO feature which does not have the smooth
character expected for the psuedo continuum produced only by Pfund
series lines running together.

Apertures with no obvious point source do not convincingly show
a CO absorption signal, though any such signal should be
weak. \citet{mey05} made predictions of the integrated light signal
from low--mass pre--main sequence (PMS) stars in massive clusters of
$\sim$ Myr age. For lower resolution spectra ($R \equiv$
$\Delta\lambda/\lambda =$ 1000), they find CO bandhead absorption
strengths of about one percent in integrated light. Our spectrum at
higher resolution shows an absorption strength at 22935 \AA \ of about
five percent. Atomic features are also weak in the \citet{mey05}
spectra (\aple 1 $\%$) and consistent with non--detections in our
spectra (that have S/N $\sim$ 50). The low--mass signal depends on the
amount of nebular emission (which degrades it) and age (younger PMS
stars are brighter and have stronger features). The fact that we
appear to detect low--mass stars in G45.45$+$0.06 suggests the cluster
is quite young. Future work on modeling the low--mass component in
UC\ion{H}{2} regions might provide more quantitative age information than has
been available to date for the associated massive stars. Precise ages
are key to understanding disk dissipation and triggering time scales
as mentioned above. 

\subsection{$T_{\rm eff}$ of the O--type stars}

The spectral type assignment made in \S~3 for sources ``m'' and ``n''
is accurate to a few sub--types. The infrared spectral type may be
associated with optical types and an effective temperature determined
for the stars.  Previous estimates of the temperature vs. spectral
type \citep{vac96} would give $T_{\rm eff}$ $\approx$ 44000 K to 48000
K (by association with the optical spectral types). Recent work using
line--blanketed atmospheres \citep{mar05} has resulted in the
effective--temperature scale ``cooling'' so that these optical
spectral types would give lower effective temperatures: $T_{\rm eff}$
$\approx$ 39000 K to 43000 K, a significant change. \citet{rep05} used
line blanketed models to derive O star effective temperatures directly
from near--infrared and optical spectra. It is found that the
effective temperatures are generally in good agreement for derivations
using near--infrared and optical spectra. \citet{rep05} report $T_{\rm
  eff}$ $=$ 42000 and 45000 derived from fits to near--infrared
spectra for an O4V star and O3V star, respectively. We thus expect the
effective temperature for sources ``m'' and ``n'' to lie within the
range given by \citet{mar05}, i.e. about $T_{\rm eff}$ $=$ 41000 K
$\pm$ 2500 K.

\begin{figure}
\plotone{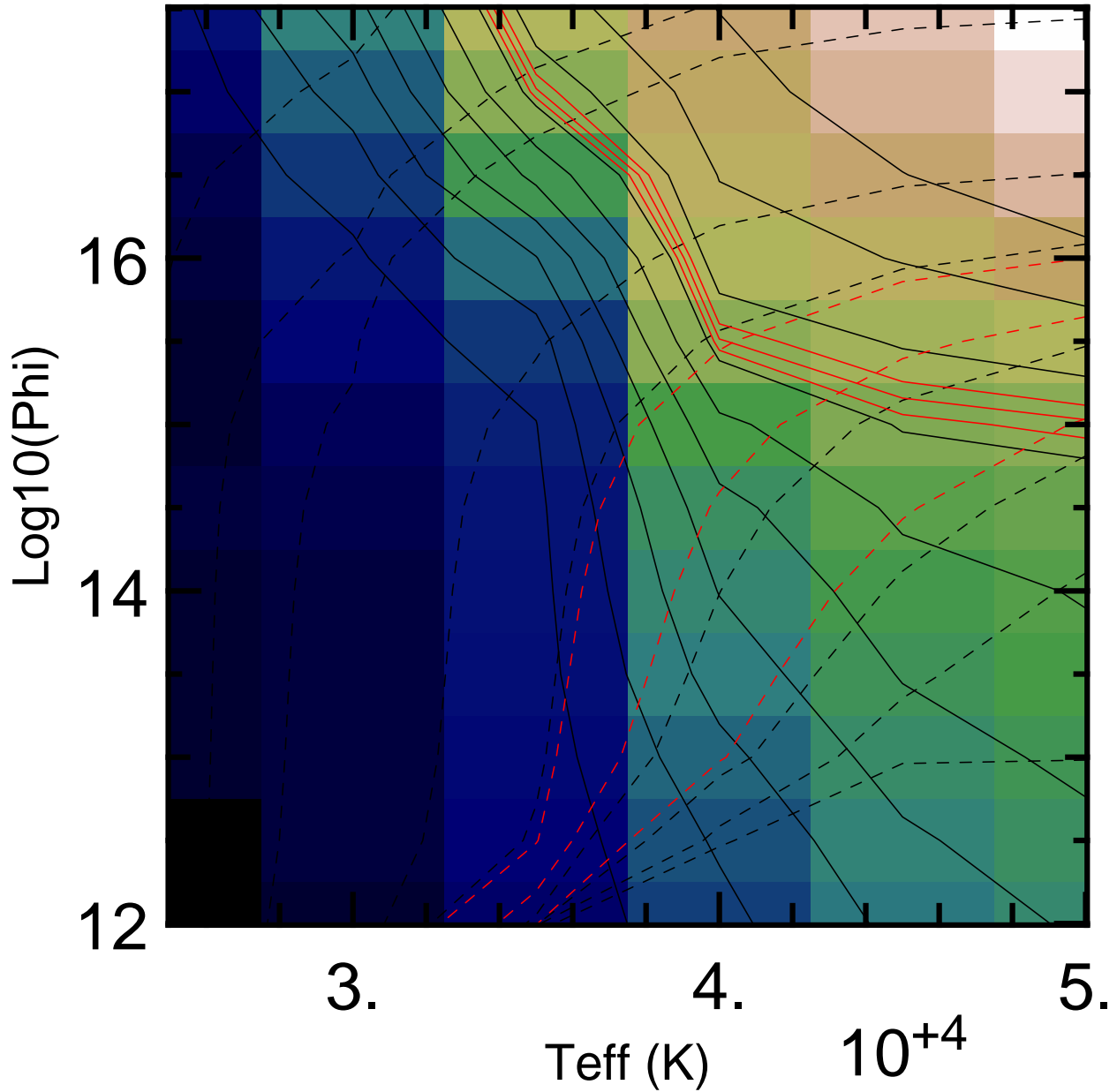}
\caption{The ratio of HeI 21126 to Br$\gamma$ (reddening free or
intrinsic) for a grid of Cloudy ionization models is shown in color
and with solid contours. The red solid contour is the observed value
(0.037, corrected for $A_K =$ 2 and shown with two other contours
corresponding to $\pm$ 0.5 mag of extinction). The dashed contours are
the results from the same models for the ratio of Br$\gamma$ (again
corrected for 2 $\pm$ 0.5 mag of extinction) to the total dust
emission. The observed dust emission is taken from the spectral energy
distribution presented by \citet{feldt98}; see text. The intersection
of the observed values for the two contour families is consistent with
the expectation of the $T_{\rm eff}$ from the infrared spectral types
of sources ``m'' and ``n''.
\label{cloudy}}
\end{figure}

The observed emission--line and dust properties of G45.45$+$0.06 can
also be used to constrain the effective temperature of exciting
source(s). We used Cloudy\footnote{Calculations were performed with
version 07.02 of Cloudy, last described in detail by \citet{fer98}}
ionization models to predict the ratio of HeI (21126 \AA) to
Br$\gamma$ and the total far infrared flux for a range of $T_{\rm
eff}$ and ionizing photon flux. The results are shown in
Figure~\ref{cloudy}. The simulations follow the procedure described by
\citet{bd99} but use HeI 21126 \AA \ instead of HeI 20587 \AA. The
simulations used the \citet{ck04} log($g$) $=$ 5 models, an electron
density of 10$^4$, and ``$\_$ism'' abundances and were run until the
zone temperature was 100 K to better simulate an UC\ion{H}{2} region. The
density is constrained by the ratio of [\ion{Fe}{3}] lines (21457 \AA/21284
\AA $=$ 0.4 corrected for $A_K$ $=$ 2; see Table~1 and
Lutz et al. 1993).

The \ion{He}{1}  to Br$\gamma$ ratio (0.034; see Table~1) is uniform
across the two pointings and as shown by \citet[see also
Benjamin et al. 1999]{han02} suggests an exciting star whose
$T_{\rm eff}$ is greater than about 40000 K. A correction of 1.08 was
used to account for the differential reddening expected for $A_K =$ 2
mag \citep{feldt98}. Figure~\ref{cloudy} shows in color and {\it
solid} contours the predicted ratio of \ion{He}{1} to Br$\gamma$ for a grid of
models whose axes are the log of incident ionizing flux and the
central source effective temperature. The observed line ratio (0.037)
is shown in a solid red contour (corrected for reddening) along with
solid contours corresponding to $\pm$ 0.5 mag of extinction. The ratio
increases monotonically at constant ionizing photon flux as the
$T_{\rm eff}$ increases. The exciting source $T_{\rm eff}$ shows a
range of values for the observed contour depending on the ionizing
photon flux.

This degeneracy may be broken, in principle, by plotting the observed
ratio of Br$\gamma$ to the total dust emission. In practice, this
comparison is difficult due to the different aperture sizes used to
map these two quantities. To make an approximate comparison, consider
the spectral energy distribution (SED) given by \citet{feldt98} using
their observations at near--infrared and mid infrared wavelengths
coupled with IRAS data points. The flux density peaks near 100
\mic. Integrating this SED and dividing by the Feldt et al. 11$''$
diameter aperture, we find a dust surface brightness of
1.3$\times$10$^{-9}$ erg cm$^{-2}$--s$^{-1}$ per square arc second
where we have reduced the IRAS flux by a factor of three so that the
IRAS points join smoothly with the \citet{feldt98} mid--infrared data
(the difference is apparently due to the different aperture sizes as
noted by Feldt et al.). From Figure~\ref{brg}, we take a typical value
of the Br$\gamma$ flux to be 3.5$\times$10$^{-14}$ erg
cm$^{-2}$--s$^{-1}$ per square arc second. The ratio of Br$\gamma$ to
dust emission from the Cloudy grid is plotted as {\it dashed} contours
in Figure~\ref{cloudy}. The observed ratio (corrected for $A_K =$ 2.0
mag) is plotted as a red dashed contour with two other red contours
corresponding to the Br$\gamma$ value for 1.5 mag and 2.5 mag $A_K$,
respectively. The intersection of the red solid and red dashed
contours is consistent with an effective temperature in the same range
we expect from the spectral types of sources ``m'' and ``n'' and a
best value at the hot end of the range of temperatures allowed by the
spectral type.

\subsection{The Distance to G45.45$+$0.06}

The NIFS photometry for sources ``m'' and ``n'' is too uncertain to
estimate a reliable distance. In addition, we have no $H$ magnitudes
from which to derive precise extinction measurements. But a rough
consistency check can be made using the extinction map generated by
\citet{feldt98} which gives $A_K$ $\approx$ 2 mag. \citet{blum00}
provide $M_K$ values for ZAMS stars as a function of spectral
type. This relation might change when accounting for the new effective
temperature scale described above, but this change actually changes
the $M_V$ of dwarf O stars very little at the high mass end
\citep[\aple 0.1 mag,][]{mar05}, so the $M_K$
should be approximately correct. Using the derived magnitudes for
``m'' and ``n', the extinction value quoted above, and $M_K$ from
\citet{blum00} appropriate to an O3 star ($-$4.8), the distance to
G45.45$+$0.06 would be about 10 kpc when averaging over the two
sources. If the O5 $M_K$ were used instead, the distance would be 7
kpc. \citet{feldt98} adopt a value of 6.6 kpc based on radio
measurements. Given the uncertainties in the photometry, extinction,
and ZAMS $M_K$ scale, this result is roughly consistent with the
\citet{feldt98} value.

\subsection{High Excitation Emission Lines from $s$--process Elements}

The two $s$--process lines ([\ion{Kr}{3}] 21986 \AA \ and [\ion{Se}{4}] 22867
\AA) were identified by \citet{din01} based on ionization potential
and the close match between the observed wavelengths in PNe and those
expected from the fine--structure energy levels arising in the two
ions. \citet{din01} built on the work of \citet{geb91} who showed the
(then) unidentified lines were not due to H$_2$ (the wavelengths did not
match) and that the lines appeared only in intermediate excitation
planetaries. \citet{geb91} observed the compact \ion{H}{2} region IRS2 in W51
along with the PNe and found no line emission at 21986 \AA \ or 22867
\AA \ in IRS2. From these observations, \citet{geb91} deduced the
ionization potential of the parent atom or ion must be in the range
30--40 eV and that the ionization potential for the ion itself should
lie at about 40--60 eV. \citet{din01} estimated the abundances of the
two species from their observed line strengths and concluded they were
in line with expectations if there were some modest enhancement due to
$s$--processing in the PNe central stars coupled with third dredge up
to the pre--planetary surface layers.

Of the 13 PNe observed by \citet{geb91} few have effective
temperatures similar to O--stars. Several are cooler, and the majority
are much hotter \citep{kj91}. The coolest (28000 K) does not exhibit
the lines, but several with temperatures in the $\sim$ 60000--80000 K
range do show the lines and with similar ratios to Br$\gamma$ as given
in Table~1 (i.e., \aple few percent). \citet{ster07} have
embarked on a large survey of PNe for which they are doing detailed
photo--ionization modeling. They report results for a set of 10 PNe
that they are using to calibrate the atomic data for Kr and Se that
are somewhat uncertain. In any case, they have observed several cooler
PNe (37000, 56000 K). Neither of these cool PNe exhibits both [\ion{Kr}{3}]
and [\ion{Se}{4}], but each exhibits one of the lines, and in similar ratio
to Br$\gamma$ as found in this work for G45.45$+$0.06
(Table~1). The abundances of Kr and Se appear to be enhanced
in many, but not all PNe with respect to the Solar value
\citep[see][for preliminary results of the large PNe survey]{ster07b}.

The presence of two relatively high--excitation lines ([\ion{Kr}{3}] 21986
\AA \ and [\ion{Se}{4}] 22867 \AA) was unexpected given that these lines have
been identified in PNe of fairly high excitation, and are not typical
of \ion{H}{2} regions. However, search of the literature shows that these
``unidentified'' lines have been detected in a number of \ion{H}{2} regions
even though the explicit association with [\ion{Kr}{3}] 
and [\ion{Se}{4}] had not been
made. \citet{lum96} suggested they may have detected both unidentified
lines in their low resolution spectrum of G45.12$+$0.13, but could not
confidently separate the 21986 \AA \ line from H$_2$. \citet{lum96}
derived an excitation temperature in G45.12$+$0.13 between 38000 K and
42000 K. \citet{okum01} studied the same UC\ion{H}{2} region, IRS2 in W51 as
did \citet{geb91}, and detected the [\ion{Se}{4}] line. It appears the
[\ion{Kr}{3}] line is also present though again blended with H$_2$ 3--2 (S3)
line (this is evident when comparing the ``excess'' level population
at the energy corresponding to an upper level of 19086K; see their
Figure 4 and Table 5). Better angular resolution may be the reason
\cite{okum01} were able to identify the line while \citet{geb91} did
not. Another possibility is differences in the detailed slit
positioning in/around IRS2.

\citet{aspin94} detected the lines in a spectrum of the object IRS~1
that is near the exciting source GGD27--IRS of both HH80 and HH81 (but
otherwise not related). \citet{aspin94} noted the possible association
of the lines with those seen by \citet{geb91}. It is not clear what
type of object might be exciting IRS 1 \citep[see][]{y87, aspin91}. The
``unidentified'' line at 21986 \AA \ has also been detected in the
extragalactic giant \ion{H}{2} region NGC 5461. \citet{pux00} note the
H$_2$ 3--2 S(3) line is blended with a line they associate with the
unidentified line cited by \citet{geb91}.

\citet{han02} presented $K-$band spectra for a sample of UC\ion{H}{2}
regions, but do not mention detection of either of these two
lines. Their spectral coverage ran out near the 22867 \AA \ line, but
one of their spectra of a ``typical'' UC\ion{H}{2} region (see their Figure
1) shows a line near 21986 \AA \ that may be due to [\ion{Kr}{3}]. 
\citet{bik05} also surveyed a number of UC\ion{H}{2} regions, but their
spectra do not cover the two line positions. \citet{bik06}
investigated deeply embedded point sources in compact \ion{H}{2} regions;
their spectra do not show evidence of either line, but it is likely
these sources are of lower effective temperature typical of early
B--type and late O--type stars. Blum and collaborators \citep{blum99,
blum00, blum01, fig02, fig05} have surveyed giant \ion{H}{2} regions using
$K-$band spectroscopy, but they also do not report detections of these
two lines; a number of their spectra also do not cover the line
regions. The lack of sufficient wavelength coverage has been due to
the trade off in instrumental spectral resolution.  Surveys for
massive stars usually tend to the blue end of the $K-$band to cover
the O--type star diagnostic lines. Most of the investigations
mentioned above have also concentrated on the point source
spectroscopy of candidate massive stars and not the overall nebular
spectrum of the regions and objects surveyed.

\begin{figure}
\plotone{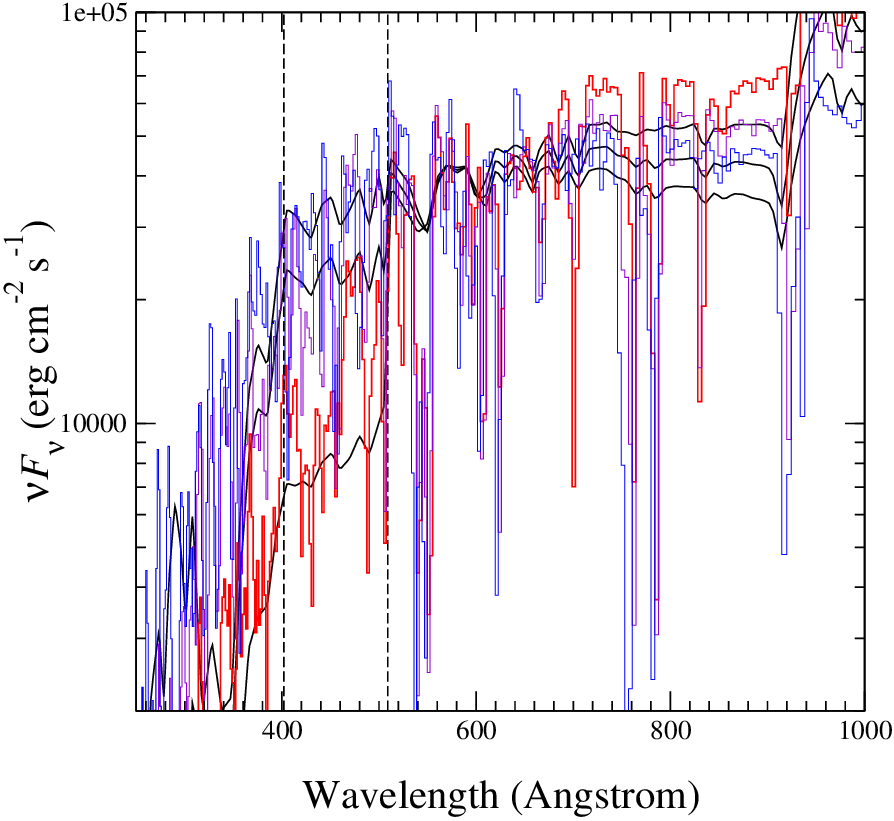}
\caption{Kurucz model atmosphere \citep[black solid lines]{ck04} and
WMBASIC \citep[colored histograms]{paul01} incident continua for
36000 K, 42000 K, and 46000 K hot stars. The vertical dashed lines
correspond to the ionization potentials of \ion{Kr}{2} and \ion{Se}{3}. The
higher $T_{\rm eff}$ models appear to have significant numbers of
ionizing photons in the spectral regions required to produce \ion{Kr}{3} 
and \ion{Se}{4}; see text. All spectra are normalized to the same total
number of ionizing photons (log(Q) cm$^{-2}$ s$^{-1}$ $=$ 15). Both
model sets show significant numbers of ionizing photons in the range
required to produce \ion{Kr}{3} and \ion{Se}{4} for the hotter stars. The cooler
stars show a precipitous drop in ionizing photons.
\label{teff}}
\end{figure}

The ionization potentials for \ion{Kr}{2} and \ion{Se}{3} are 24.4 eV
and 30.8 eV, respectively \citep{din01}. This is well within the range
of the ionizing continuum of the hotter O--type stars, particularly
for \ion{Kr}{2}. Figure~\ref{teff} shows the ionizing continua of hot
stars for a range of effective temperatures and two sets of
models. Plane parallel, hydrostatic models are plotted \citep{ck04}
along with line blanketed wind models \citep{paul01}. Both show
significant numbers of ionizing photons for the hotter stars and a
rapid decrease in ionizing photons in the range appropriate to form
\ion{Kr}{3} and \ion{Se}{4} as the effective temperature
falls. Wavelengths corresponding to the ionization potentials of
\ion{Kr}{2} and \ion{Se}{3} are shown as vertical dashed lines.

While the presence of these two lines is not ubiquitous in \ion{H}{2}
region spectra, it seems that the hotter O stars, in dense
UC\ion{H}{2} regions, can produce these lines. The distribution of the
emission within the UC\ion{H}{2} follows the ionized gas as shown in
the maps of Figure~\ref{brg} and \ref{krsel}. Our unpublished NIFS
data for the UC\ion{H}{2} region G5.89$-$0.39 show strong emission at
21986 \AA \ and possibly weaker emission at the [\ion{Se}{4}] position toward
the center of the UC\ion{H}{2} region. We were unable to identify the
exciting source in this case, but the excitation as given by the HeI
to Br$\gamma$ ratio and Br$\gamma$ to dust ratio appears similar to
that for G45.45$+$0.06. A detailed analysis of the G5.89$-$0.39 data
is in progress. Given the ionization potentials of the parent ions to
the species responsible for these lines and the strong gradient in the
ionizing continuum hardness of the O stars, we believe that these
lines will only be produced in dense \ion{H}{2} regions with the hottest
ionizing stars.

\section{Summary}

NIFS 3-dimensional imaging spectroscopy is presented for two pointings
toward the UC\ion{H}{2} region G45.45$+$0.06. The central ionizing
sources of G45.45$+$0.06 have been identified as massive O--type stars
through comparison of their $K-$band spectra to template O star
spectra. The \citet{feldt98} sources ``m'' and ``n'' are found to be
of O spectral types earlier than about kO6. In addition, two other
\citet{feldt98} sources, ``k'' and ``o'', exhibit CO bandhead emission
near 23000 \AA \ indicating that they are massive young stellar
objects. In analogy to other similar objects in other \ion{H}{2}
regions, these two massive young stellar objects likely have
significant $K-$ band excesses and thus have central stars consistent
with late O and early B stars. The remaining \citet{feldt98} sources
in the NIFS fields (``l'', ``g'', ``h'', and ``i'') show strong
compact nebular emission and rising red continua suggesting that they
too are massive young stellar objects.

The NIFS images reveal complex nebular structure and precise emission
line maps. The morphology of the UC\ion{H}{2} region suggests that the
central hot stars have triggered star formation in the material to the
north west as represented by point sources ``g'', ``h'', ``i'', and
``k'' and also to the south in the form of source ``o''. However,
there is no way to precisely assign ages to these sources relative to
the central ionizing stars that are presumably on, or near, the zero
age main sequence. Modeling the weak signal from low--mass pre--main
sequence stars may provide a much needed clock for high--mass
starforming regions. A low--mass signal is present in the NIFS data
for G45.45$+$0.06, though it is not clear if it is associated with one
or a few low--mass objects. An alternate scenario is that the massive
stars form in the cluster center and lower mass stars form around them
at the same time.

Lines of doubly ionized Kr and triply ionized Se are detected in the
NIFS spectra. Line maps show that these lines follow the ionized gas
in the UC\ion{H}{2} region as traced by Br$\gamma$ and HeI. The lines
are not common in \ion{H}{2} regions, but have been identified in
planetary nebulae (PNe) where the central stars are typically much
hotter. In fact, one or the other or both of the lines, while not
previously identified with \ion{Kr}{3} and \ion{Se}{4}, have been
detected in a number of compact \ion{H}{2} regions. The \ion{Kr}{3} line,
in particular, is blended with H$_2$ 3--2 S(3) at lower
spectral resolution making its identification more difficult.

These are $s$--process lines that show enhanced abundances over the
Solar value for some, but not all, PNe. We suggest that it may be only
the \ion{H}{2} regions excited by the hottest O stars and with high
density that can produce these lines. If so, the lines may be a useful
new diagnostic on the effective temperature of ionizing stars in
\ion{H}{2} regions.  Much of the observational data in the literature
are incomplete with respect to wavelength coverage for the detection
of these lines in a wide range of \ion{H}{2} regions, and the atomic
parameters of the ions of Kr and Se are not yet well established, but
progress is being made on both fronts.

We acknowledge the helpful comments of an anonymous referee.
The plots and analysis in this article made use of the Yorick
programing language. This article made use of the SIMBAD database at
the CDS. The authors would like to thank Tom Geballe for initially
pointing out the line identifications for Kr and Se and Dick Shaw for
useful discussions regarding planetary nebulae.



\begin{deluxetable}{lccc}
\label{lines}
\tablecaption{{\it Emission Line Identifications and Ratios}}
\tablehead{
\colhead{Line Identification\tablenotemark{a}} &
\colhead{Wavelength (\AA)\tablenotemark{b}} &
\colhead{Pointing 1 Ratio to Br$\gamma$\tablenotemark{c}} &
\colhead{Pointing 2 Ratio to Br$\gamma$\tablenotemark{c}} 
}
\startdata
HeI $^1$S--$^1$P$^{\circ}$ & 20587	& 0.865 $\pm$  0.0015 &  0.812 $\pm$  0.0011 \\
HeI $^3$P$^{\circ}$--$^3$S & 21126	& 0.035 $\pm$  0.0003 &  0.034 $\pm$  0.0002 \\
HeI $^1$P$^{\circ}$--$^1$S & 21138	& 0.012 $\pm$  0.0002 &  0.009 $\pm$  0.0001 \\
H$_2$ v$=$1--0 S(1)	& 21218	& 0.015 $\pm$  0.0002 &  0.015 $\pm$  0.0001 \\
$[$\ion{Fe}{3}$]$ $^3$H--$^3$G & 21457	& 0.007 $\pm$  0.0002 &  0.003 $\pm$  0.0001 \\ 
HeI $^3$D--$^3$F$^{\circ}$ & 21614	& 0.022 $\pm$  0.0002 &  0.020 $\pm$  0.0001 \\
HeI $^1$D--$^1$F$^{\circ}$ & 21623	& 0.006 $\pm$  0.0002 &  0.007 $\pm$  0.0001 \\
HeI     & 21646	& 0.044 $\pm$  0.0003 &  0.049 $\pm$  0.0002 \\
Br$\gamma$ (HI 7--4)	& 21661	& 1.000 $\pm$  0.0017 &  1.000 $\pm$  0.0012 \\
HeI?\tablenotemark{d}	& 21820	& 0.005 $\pm$  0.0002 &  0.004 $\pm$  0.0001 \\
HeI\tablenotemark{d} $^1$P$^{\circ}$--$^1$D & 21846	& 0.006 $\pm$  0.0002 &  0.003 $\pm$  0.0001 \\
$[$\ion{Kr}{3}$]$ $^3$P$_1$--$^3$P$_2$ & 21987& 0.008 $\pm$  0.0002 &  0.006 $\pm$  0.0001 \\
$[$\ion{Fe}{3}$]$ $^3$H--$^3$G& 22184	& 0.018 $\pm$  0.0002 &  0.017 $\pm$  0.0001 \\
H$_2$ v$=$1--0 S(0)	& 22233	& 0.006 $\pm$  0.0002 &  0.005 $\pm$  0.0001 \\
$[$\ion{Fe}{3}$]$ $^3$H--$^3$G & 22427 & 0.006 $\pm$  0.0002 &  0.006 $\pm$  0.0001 \\
H$_2$ v$=$ 2--1 S(1)& 22477	& 0.004 $\pm$  0.0002 &  0.001 $\pm$  0.0001 \\
$[$\ion{Se}{4}$]$ $^2$P$_{3/2}$--$^2$P$_{1/2}$ & 22867	& 0.013 $\pm$  0.0002 &  0.012 $\pm$  0.0001 

\enddata

\tablenotetext{a}{Emission line identifications were made using the
on--line data base at NIST \citep{nist}. For HeI, see
\citet{dm98}. \ion{Fe}{3} lines were obtained from Peter van Hoof's web page
(http://www.pa.uky.edu/~peter/atomic/) that use energy levels from
the NIST database. H$_2$ lines were taken from \citet{bd87}.}

\tablenotetext{b}{Vacuum}

\tablenotetext{c}{Emission lines were extracted from a 2$''$ diameter
aperture near center of field of view. No local sky aperture was used
as in the case for point source extraction. See text.}

\tablenotetext{d} {\citet{bik05} identify this line with HeI.}

\end{deluxetable}
\end{document}